\documentclass{pasj00}
\begin{document}
\SetRunningHead{A. Bamba et al.}{A Diffuse Source AX~J1843.8$-$0352}
\Received{2001 March 29}%{yyyy/mm/dd}
\Accepted{2001 June 8}%{yyyy/mm/dd}

\title{A Diffuse X-Ray Source, AX~J1843.8$-$0352: 
Association with the Radio Complex G28.6$-$0.1 
and Identification of a New Supernova Remnant}

%%% begin:list of authors
\author{%
Aya \textsc{Bamba}\altaffilmark{1}
Masaru \textsc{Ueno}\altaffilmark{1}
Katsuji \textsc{Koyama}\altaffilmark{1, 3}
and
Shigeo {\textsc Yamauchi}\altaffilmark{2}}

\altaffiltext{1}
{Department of Physics, Graduate School of Science, Kyoto University, 
Sakyo-ku, Kyoto, 606-8502}
\email{bamba@cr.scphys.kyoto-u.ac.jp, masaru@cr.scphys.kyoto-u.ac.jp, koyama@cr.scphys.kyoto-u.ac.jp}
\altaffiltext{2}
{Faculty of Humanities and Social Sciences, Iwate University, 
3-18-34 Ueda, Morioka, Iwate 020-8550}
\email{yamauchi@hiryu.hss.iwate-u.ac.jp}
\altaffiltext{3}{CREST, Japan Science and Technology Corporation (JST), 
4-1-8 Honmachi, Kawaguchi, Saitama 332-0012}
%%% end:list of authors

%%% Please use the following style in case that sorting by 
%%% affilation is impossible. 
%
% \author{%
%   D-Firstname \textsc{D-Familyname}\altaffilmark{1}
%   E-Firstname \textsc{E-Familyname}\altaffilmark{1,2}
%   and
%   F-Firstname \textsc{F-Familyname}\altaffilmark{2}}
% \altaffiltext{1}{Address of Institute}
% \email{ddddd@xxx.xxx.xx.xx}
% \email{eeeee@xxx.xxx.xx.xx}
% \altaffiltext{2}{Address of Institute}

%% `\KeyWords{}' always has to be placed before `\maketitle'.
 %Do NOT move this preamble from here!

\maketitle

\begin{abstract}
ASCA discovered an extended source in the Scutum constellation.
The X-ray morphology is an elliptical shape elongated from north to south
with a mean diameter of about 10$^\prime$.  
The image center is located at
$\rm RA_{2000}$ = \timeform{18h43m53s}, 
$\rm DEC_{2000}$ = $-\timeform{03D52'55''}$
(hereafter, AX~J1843.8$-$0352).
The north and south rims of AX~J1843.8$-$0352 are associated with
non-thermal radio sources C and F of the G28.6$-$0.1 complex
(Helfand et al.\ 1989).
The X-ray spectrum was fitted with a model of either a thin thermal plasma  
in non-equilibrium ionization
of a temperature 5.4~keV or a power-law of photon index $2.1$.
The absorption column is (2.4--4.0)~$\times~10^{22}$~cm$^{-2}$,
which probably places this source in the Scutum arm.
With a reasonable distance assumption of about 7~kpc,
we estimate the mean diameter and X-ray luminosity to be $\sim$~20~pc
and $\sim~3~\times ~10^{34}$~erg~s$^{-1}$, respectively.
Although a Sedov solution for a thin thermal plasma model gives
parameters of a young shell-like SNR,
no strong emission lines are found with the metal abundances being
$\sim$ solar.   
Thus, a more likely scenario for both the radio and X-ray spectra and 
the morphologies is that  AX~J1843.8$-$0352 is a shell-like SNR
which predominantly emits synchrotron X-rays.
\end{abstract}

\section{Introduction}

The most complete catalogue of the galactic supernova remnants (SNRs)
is found in the radio band; about 220 radio SNRs are currently 
cataloged \citep{green}.  In the X-ray band  however,  the cataloged number 
is less than half of the radio SNRs.  
On the other hand, recent surveys with ROSAT and ASCA
found new X-ray SNRs \citep{koyama1997},
which were later identified with radio-faint SNRs.
We thus suspect that new SNRs which have escaped radio detection
can be found in the X-ray band. 
Of particular interest are synchrotron X-rays from the shells of SNRs
(e.g., SN~1006; \cite{koyama1995}),
where electrons should be accelerated to high energies of
up to about 1~TeV or even more,
possibly with the first-order of the Fermi acceleration,
which implies that the shell of SNRs can be the birthplace of cosmic rays. 
The gain (acceleration) rate of electrons is proportional
to the magnetic field strength ($B$),
while the synchrotron energy loss of electrons is proportional to $B^2$.
Therefore, high-energy electrons responsible for the synchrotron X-rays are
likely to exist in a shell with a rather weak $B$,
where the radio flux should be faint,
because the flux is proportional to $B^2$.

With these ideas in mind,
we have searched for diffuse hard X-ray sources in the galactic plane
from the ASCA survey data, and found a handful of candidates. 
For some of them, we have carried out long exposure observations
to investigate whether these sources are new SNRs
which emit synchrotron X-rays.
This paper reports on a survey and pointing observations of
the SNR candidate AX~J1843.8$-$0352,
and discusses the X-ray emission mechanisms of this source. 

\section{Observations and Data Reduction}

ASCA observations near the galactic Scutum arm
($l$=\timeform{28D}--\timeform{29D})
were made on 1993 October 22 (obs.\ 1)
and on 1997 April 22 (obs.\ 2)
as a part of the galactic plane survey project.
We then proceeded to a follow-up observation with deeper exposure
on 1999 May 23 (obs.\ 3).
ASCA carries four XRTs (X-Ray Telescopes, \cite{serlemitsos})
with two GISs (Gas Imaging Spectrometers, \cite{ohashi}) and 
two SISs (Solid-state Imaging Spectrometers, \cite{burke})
on the focal planes.
Since our target is a diffuse  source 
with a size comparable to or larger than the SIS field of view,
we do not refer to the SIS in this paper.
In all of the observations, the GISs were operated in the nominal PH mode.
We rejected the GIS data obtained in the South Atlantic Anomaly,
in low cut-off rigidity regions ($<6$~GV), 
or when the target's elevation angle was low ($< 5^\circ$).
Particle events were removed by the rise-time discrimination method
\citep{ohashi}.
After screening, the total available exposure times of obs.\ 1, 2, and 3
were $\sim 19$~ks,  $\sim 10$~ks, and $\sim 42$~ks, respectively.
To increase the statistics, the data of the 3 observations 
and those of two detectors, GIS-2 and GIS-3, were all combined.

\section{Analysis and Results}

\begin{figure}
  \begin{center}
    \FigureFile(80mm,80mm){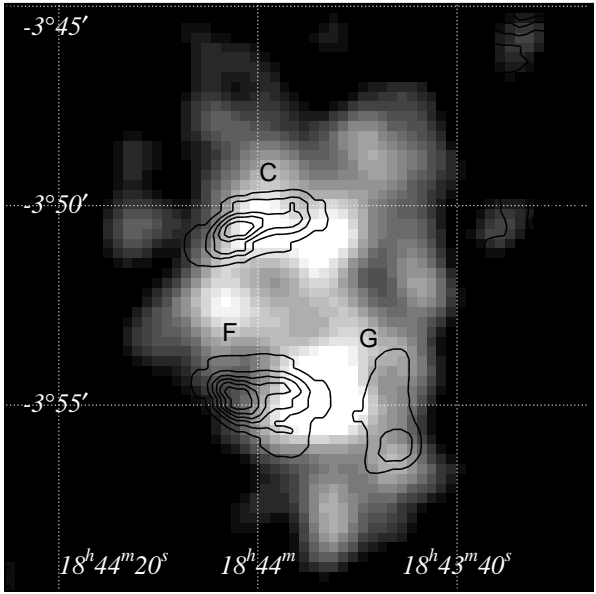}
    %%% \FigureFile(width,height){filename}
  \end{center}
  \caption{ASCA GIS image in the 0.7--7.0~keV band  
overlaid on the NVSS 20~cm radio contours. (This map was
retrieved from the web site of the NRAO VLA Sky Survey, 
http://www.cv.nrao.edu/\~{}jcondon/nvss.html.)
Radio sources are labeled following \citet{helfand}.}\label{fig1}
\end{figure}

The GIS image in the 0.7--7.0~keV band near the Scutum region is given 
in figure~\ref{fig1}.
We can see diffuse X-rays with an elliptical shape of its center at 
$\rm (RA, DEC)_{J2000}$ = (\timeform{18h43m53s}, $-\timeform{03D52'55''}$).
We hence designate this source as AX~J1843.8$-$0352.

\begin{figure}
  \begin{center}
    \FigureFile(80mm,80mm){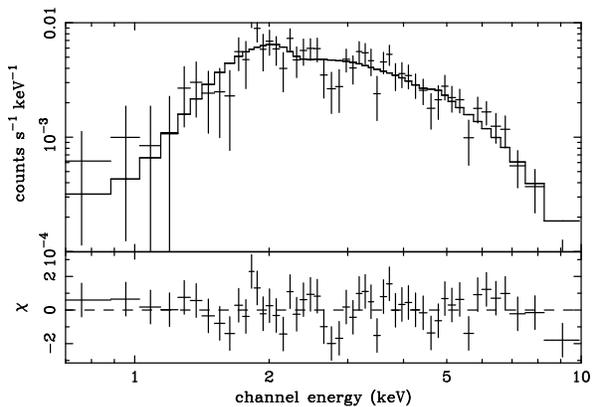}
    %%% \FigureFile(width,height){filename}
  \end{center}
  \caption{Background-subtracted spectrum of the combined GIS~2 and 3 
data of the 3 observations. The solid line is the best-fit power-law model 
(upper panel). Lower panel shows the data residuals from the best-fit model.}
\label{fig2}
\end{figure}

The X-ray spectrum is made using the photons in an ellipse area of
\timeform{13'} (major axis) by \timeform{9'} (minor axis),
while the background spectrum is obtained
from an elliptical ring around the source.
The background-subtracted spectrum is given in figure~\ref{fig2},
where the total photons in the 0.7--10~keV band are 
$\rm 2.7 \times 10^{3}$ counts.
We then fit the spectrum with two models of a power-law and 
a thin thermal plasma  
in non-equilibrium ionization (NEI plasma; coded by \cite{masai}).
The NEI condition is characterized by the ionization parameter, $n_{\rm e}t$,
where $n_{\rm e}$ and $t$ are the electron density and elapsed time
after the plasma has been heated-up.
The collisional ionization equilibrium is achieved when $\log (n_{\rm e}t)$
becomes larger than  $\sim 12$.    
The absorption column is calculated using the cross sections of
\citet{anders} with solar abundances.
These two models are statistically acceptable
with the best-fit parameters given in table~\ref{table1}.
In figure~\ref{fig2}, we show the best-fit power-law model 
(the solid line).

\section{Discussion}

The best-fit $N_{\rm H}$ value of (2.4--4.0)~$\times 10^{22}$~cm$^{-2}$
is similar to the nearby  SNRs,
Kes~73 [(1.6--2.3)~$\times 10^{22}$~cm$^{-2}$; \cite{gotthelf}] and
Kes~75 [$(2.9\pm 0.4)\times 10^{22}$~cm$^{-2}$; \cite{blanton}],
both are located near the same galactic coordinate as AX~J1843.8$-$0352,
at (\timeform{27.4D}, \timeform{0.0D}) and 
(\timeform{29.7D}, $-\timeform{0.3D}$), respectively.  
Using the H~$_{\rm I}$ absorption data, the distances of Kes~73 and Kes~75 are
estimated to be 6--7.5~kpc \citep{sanbonmatsu}
and 9--21~kpc (\cite{caswell}; \cite{becker1984}), respectively.  
\citet{koyama1990} found many transient sources at the tangent 
point of the Scutum arm (about 8.5~kpc) with absorptions of around 
10$^{23}$~cm$^{-2}$,
which are larger than that of AX~J1843.8$-$0352.  
Accordingly, we infer that the distance of AX~J1843.8$-$0352 is in between 
6--8.5~kpc.
Hereafter, we assume the distance to be 7~kpc. 
The X-ray luminosity (0.1--10.0~keV) and the source size (the mean diameter)
are then  estimated to be $3\times 10^{34}$~erg~s$^{-1}$ and 20~pc,
respectively.  These parameters are similar to those
of shell-like young SNR.

In the survey data of 6~cm and 20~cm bands 
(\cite{becker1994}; \cite{reich}), we find 
some sources near  AX~J1843.8$-$0352.
These radio sources, however, are not cataloged as SNRs \citep{green}.
\citet{helfand} resolved the radio complex called G28.6$-$0.1
into individual radio sources, A--I, with the VLA 20~cm band 
observations. Combining with the 6~cm band data by \citet{altenhoff},
the spectral indices of C and F were estimated to be 0.5--0.6,
or non-thermal radio sources; the morphology of F, 
at least, was a  cluster head-tail source, although 
possibility a part of a galactic SNR was not excluded. 
Our present observation strongly suggest that AX~J1843.8$-$0352
is an X-ray SNR and the non-thermal radio 
sources C and F are the radio counterparts.   
Because source G has a more flat radio spectrum,
it would be thermal, possibly an H~$_{\rm II}$ region and an unrelated source,
although located inside of AX~J1843.8$-$0352. 

The flux  densities  of the  radio complex G28.6$-$0.1 at 
the 6~cm  and 20~cm bands 
are 0.9~Jy  and 2.1~Jy, respectively (\cite{altenhoff}; \cite{helfand}). 
Extrapolating these values, 
we estimate the surface brightness of AX~J1843.8$-$0532 to be 
$\rm 9.7\times 10^{-22}\ W\ m^{-2}Hz^{-1}sr^{-1}$ at 1~GHz (in 30~cm band).
This value is lower than those of the typical galactic SNRs
with a diameter $\sim$~20~pc,
and is similar to that of SN~1006 (\cite{case}; \cite{winkler}).

Assuming a thin thermal NEI plasma model,
and applying a Sedov model to AX~J1843.8$-$0352, the ambient density, age,  
explosion energy, and total mass are estimated to
be 0.2~cm$^{-3}$,  
2700~yr, 0.9$\times$10$^{51}$~erg, and 20~$M_\odot$, respectively.   
If the remnant is the result of a core collapse explosion,
20~$M_\odot$  could be comparable
to the original mass of the progenitor,
hence the remnant would be in between the free-expansion and Sedov phases. 
Accordingly, the age and density inferred with the Sedov assumption
should be regarded as upper limits. 

For a Type Ia explosion, like SN~1006, the 20~$M_\odot$ is significantly
larger than the ejecta mass, and is hence in the Sedov phase.
This remnant (at 7~kpc) has a similar physical size to
that of SN~1006 (\timeform{30'} at 2~kpc).
However, the X-ray luminosity is about an order of magnitude less;
hence, the density and possibly the age would be smaller than those of SN~1006.

Thus, in any case (whether core collapsed or type Ia SNRs),
the real age and ambient density of AX~J1843.8$-$0532
are significantly smaller  than the inferred values of 2700~yr 
and 0.2~cm$^{-3}$, respectively.
The best-fit ionization parameter of
$\log (n_{\rm e}t)$ is $\sim 10$,
which also supports a smaller age and density.

For a core collapse young SNR, the abundances of  Si and S would be enhanced,
because the ratio of the stellar mass to the swept-up mass is high.
For a Type Ia, Fe  should be enhanced, because
iron is the most dominant element in the ejected material.
No strong lines from these elements are, however found,
and the abundances are nearly one  solar.
We hence conclude that the thin thermal plasma scenario
is unlikely for this young SNR. 
The best-fit NEI temperature of 5.4~keV is higher than any other 
young SNRs, such as Cas A, Tycho, Kepler, and W~49B
[2.56$\pm$0.05~keV \citep{bleeker},
2.3$\pm$0.3~keV \citep{decourchelle},
$3.1^{+0.5}_{-0.4}$~keV \citep{kinugasa}, and
2.2--2.7~keV \citep{hwang}, respectively], 
but is similar to the best-fit thermal model temperature 
for SN~1006 and G347.3$-$0.5, the typical SNRs of non-thermal X-rays
[7--10~keV \citep{ozaki} and 3.8$\pm$0.3~keV \citep{koyama1997}, respectively].
Thus, by analogy, we expect the emission from this remnant to be
also non-thermal.  
In fact, a power-law model with a photon index of 2.1 (an energy index of 1.1)
also fit well to the X-ray spectrum of AX~J1843.8$-$0352.
The combined radio spectrum of sources C and F
is smoothly connected to the 
X-ray spectrum of AX~J1843.8$-$0352
with a power-index break between the radio and X-ray bands. 
The power-law index of about 0.5 in the radio band is explained by
synchrotron emissions of high-energy electrons having a power index of 2.0.  
The larger power-law energy index of 1.1 (photon index = 2.1) in the X-ray 
band should be due to the synchrotron energy loss;
higher energy electrons loose their energy more rapidly
than lower energy electrons. 

Thus, the spectra and morphologies in both the radio and the X-ray bands 
are fully consistent with the scenario
that AX~J1843.8$-$0352 is a shell-like SNR with synchrotron X-rays,
following well-established examples: SN~1006 \citep{koyama1995}, 
G347.3$-$0.5 \citep{koyama1997} and RX~J0852.0$-$4622 
\citep{allen}. 

The physical size of AX~J1843.8$-$0352 is comparable to that of SN~1006;
still, the X-ray luminosity is lower than that of SN~1006.
This may lead to a lower magnetic field strength ($B$) and/or
a smaller electron number density ($n_{\rm e}$),
possibly due to a lower density of the inter-stellar medium (ISM).
In the low-density ISM,
the forward shock velocity remains high for a sufficiently long enough time  
to produce TeV electrons, which emit synchrotron (non-thermal) X-rays.  
The lower is $n_{\rm e}$,
the lower is the thermal X-ray flux from the forward shock.
Also, a strong reverse shock might not form,
because the velocity difference between the ejecta
and the shock may not be dramatic; hence, thermal X-rays would be suppressed.

For further quantitative studies of the synchrotron scenario,
we need a more detailed comparison of the X-ray spectrum and morphology with 
those of the radio band, which can be achieved
with future observations of Chandra and/or XMM-Newton,
which have better spatial resolution and a larger effective area
than ASCA.

\section{Summary}

1. We found a diffuse hard X-ray source, AX~J1843.8$-$0352, associated with 
the non-thermal radio complex G28.6$-$0.1 in the galactic Scutum arm.\\
2. The X-ray spectrum is fitted with either a high-temperature 
(5.4~keV) NEI plasma model or a non-thermal power-law model of index 2.1.\\
3. The $N_{\rm H}$ absorption is (2.4--4.0)~$\times 10^{22}$~cm$^{-2}$,
which constrains the source distance to be 6--8.5~kpc.\\
4. The morphology (20~pc diameter sphere), association with 
the non-thermal radio sources, X-ray spectrum and luminosity strongly 
support that AX~J1843.8$-$0352 is a newly identified  SNR.\\
5. A high-temperature thin thermal plasma for this SNR is unlikely,
and a non-thermal process from the shell is a more plausible scenario; 
AX~J1843.8$-$0352 predominantly emits synchrotron X-rays from the shells. 

\vspace{1pc}\par

The authors are grateful to all members of the ASCA team.
Our particular thanks are due to the ASCA galactic plane survey team.
We also thank an anonymous referee for his/her insightful suggestions and comments.
A.B. is supported by JSPS Research Fellowship for Young Scientists.

\begin{longtable}{lccccc}
  \caption{Best-fit parameters of AX~J1843.8$-$0352
for a power-law, a thin thermal plasma in non-equilibrium ionization (NEI) 
models $^\ast$.}\label{table1}
   \hline\hline
 Model& $\Gamma$ / $kT$&$N_{\rm H}$ &Abundance$^{^\dagger}$ & log $n_{\rm e}t$ &$\chi ^{2}$\\
 &  .../(keV) &($10^{22}$~cm$^{-2}$)& & s cm$^{-3}$& /degrees of freedom
\endfirsthead
\endhead
\hline
Power-law & $2.1^{+0.3}_{-0.4}$&$2.6^{+0.8}_{-0.6}$& ... & ... & 45.4/47 \\
NEI plasma & $5.4^{+3.4}_{-1.8}$&$4.0^{+1.8}_{-1.2}$&$1.5^{+1.9}_{-1.2}$&$9.6^{+0.3}_{-0.3}$ &  32.5/46 \\
\hline
\multicolumn{5}{l}
{$^\ast$ Errors and upper-limits are at the 90\% confidence for one relevant parameter.}\\
\multicolumn{5}{l}{$^\dagger$ Assuming the solar abundance ratio
\citep{anders}.}\\
\end{longtable}

\end{document}